\documentclass[amsfonts,12pt,thmsa,sw20bams]{article}

\usepackage{graphicx}
\usepackage{amssymb}

\input tcilatex
\begin{document}

\author{Choulakian Vartan and Abou Samra Ghassan \and Universit\'{e} de
Moncton, Canada, \and email: vartan.choulakian@umoncton.ca;
ghassan.abou.samra@umoncton.ca}
\title{Mean absolute deviations about the mean, the cut norm and taxicab
correspondence analysis}
\date{January 2020}
\maketitle

\begin{abstract}
Optimization has two faces, minimization of a loss function or maximization
of a gain function. We show that the mean absolute deviations about the
mean, $d$, maximizes a gain function based on the power set of the
individuals, and it is equal to twice the value of its cut-norm. This
property is generalized to double-centered and triple-centered data sets.
Furthermore, we show that among the three well known dispersion measures,
standard deviation, least absolute deviation and $d$, $d$ is the most robust
based on the relative contribution criterion. More importantly, we show that
the computation of each principal dimension of taxicab correspondence
analysis corresponds to balanced 2-blocks seriation. Examples are provided.

Key words: Mean absolute deviations about the mean; cut norm; balanced
2-blocks seriation; taxicab correspondence analysis.
\end{abstract}

\section{\bf Introduction}

Optimization has two faces, minimization of a loss function or maximization
of a gain function. The following two well known dispersion measures, the
variance ($s^{2})$ and mean absolute deviations about the median $(LAD$),
are optimal because each minimizes a different loss function
\begin{eqnarray}
s^{2} &=&\frac{\sum_{i=1}^{n}(y_{i}-\overline{y})^{2}}{n}  \nonumber \\
&\leq &\frac{\sum_{i=1}^{n}(y_{i}-c)^{2}}{n}
\end{eqnarray}%
and
\begin{eqnarray}
LAD &=&\frac{\sum_{i=1}^{n}|y_{i}-median|}{n}  \nonumber \\
&\leq &\frac{\sum_{i=1}^{n}|y_{i}-c|}{n},
\end{eqnarray}%
where $y_{1},\ y_{2,}...,y_{n}\ $and $c$ represent a sample of ($n+1)$
values. To our knowledge, no optimality property is known for the mean
absolute deviations about the mean defined by
\begin{equation}
d=\frac{\sum_{i=1}^{n}|y_{i}-\overline{y}|}{n},
\end{equation}%
even though it has been studied in several papers for modeling purposes by,
see among others, Pham-Gia and Hung (2001), Gorard (2015), Yitzhaki and
Lambert (2013). Pham-Gia and Hung (2001) and Gorard (2015) essentially
compare the dispersion measures $d$ and $s$ in the statistical litterature,
with their preference clearly oriented towards $d$ for its simple
interpretability. While Yitzhaki and Lambert (2013) compare the statistics $d
$ and $LAD$ with the Gini dispersion measure and conclude that
\textquotedblright The downside of using ($d$ and $LAD$) is that robustness
is achieved by omitting the information on the intra-group
variability\textquotedblright .

The following inequality $LAD\leq d\leq s$ is well known: it follows from
(2) and the fact that $ns^{2}=\sum_{i=1}^{n}w_{i}^{2}-n\overline{w}^{2}\geq
0 $, where $w_{i}=|y_{i}-\overline{y}|.$

$d$ is the measure of dispersion used in taxicab correspondence analysis
(TCA), an L$_{1}$ variant of correspondence analysis (CA), see Choulakian
(2006). An explanation for the robustness of $d$ is the boundedness of the
relative contribution of a point, see Choulakian (2008a, 2008b, 2017) and
Choulakian et al. (2013a, 2013b, 2014). However, this paper provides further
details on $d$, relating it to cut-norm and balanced 2-blocks seriation for
double-centered data. Choulakian (2017) argued that often sparse contingency
tables are better visualized by TCA; here, we present an analysis of a 0-1
affinity matrix, where TCA produces a much more interpretable map than CA.
We see that repetition and experience play an indispensable and illuminating
role in data analysis.

This paper is organized as follows: In section 2, we show the optimality of
the $d,\ s^{2}$ and $LAD$ statistics based on maximizing gain functions, but
$d$ beats $s^{2}$ and $LAD$ with respect to the property of relative
contribution of a point (a robustness measure used in french data analysis
circles based on geometry): this results from Lemma 1, which states the fact
that for a centered vector $d$ equals twice its cut-norm; sections 3 and 4
generalize the optimality result of the $d$ to double-centered and
triple-centered arrays; and we conclude in section 5. Balanced 2-blocks
seriation of a matrix with application to TCA is discussed in section 3.

\section{Optimality of d}

We construct the centered vector ${\bf x}={\bf y}-\overline{y}\ {\bf 1}_{n}$%
, where ${\bf 1}_{n}$ is composed of $n$ ones. Let $I=\left\{
1,2,...,n\right\} $ and $I=S\cup \overline{S}$ a binary partition of $I$. We
have%
\begin{eqnarray*}
\sum_{i\in I}x_{i} &=&0 \\
&=&\sum_{i\in S}x_{i}+\sum_{i\in \overline{S}}x_{i};
\end{eqnarray*}%
from which we deduce%
\begin{equation}
\sum_{i\in S}x_{i}=-\sum_{i\in \overline{S}}x_{i}.
\end{equation}%
We define the cut-norm of a centered vector {\bf x} to be $||{\bf x||}%
_{\boxdot}=\max_{S}\sum_{i\in S}x_{i}=\sum_{i\in S_{opt}}x_{i}$, where $%
S_{opt}=\left\{ i:x_{i}\geq 0\text{ for }i=1,2,...,n\right\} .$ By casting
the computation of $d$ as a combinatorial maximization problem, we have the
following main result describing the optimality of the $d$-statistic over
all elements of the power set of $I$.\bigskip

{\bf Lemma 1}: (2-equal parts property) $d=2||{\bf x||}_{\boxdot }\geq
2\sum_{i\in S}x_{i}/n$\ \ \ \ for \ all\ \ $S\subset I$.\medskip

Proof: Easily shown by using (4).

\bigskip

{\bf Corollary 1}: $d\geq {\bf x}^{\prime }{\bf u/}n$\ \ \ for\ \ ${\bf u}%
\in \left\{ -1,1\right\} ^{n}.\bigskip $

Proof: By defining ${\bf u}_{opt}(i)=1$ if $i\in S_{opt}$ and$\ {\bf u}%
_{opt}(i)=-1$ if $i\in \overline{S}_{opt}$, we get $d={\bf x}^{\prime }{\bf u%
}_{opt}{\bf /}n\geq {\bf x}^{\prime }{\bf u/}n.$\bigskip

{\bf Corollary 2:} $LAD\geq ({\bf y}-median$ ${\bf 1}_{n})^{\prime }{\bf u/}%
n $\ \ \ for\ \ ${\bf u}\in \left\{ -1,1\right\} ^{n}.$\bigskip

Corollary 2 shows that LAD has a second optimality property. We emphasize
the fact that the optimizing function in (2) is a univariate loss function
of $c\in
\mathbb{R}
$; while the optimizing function in Corollary 2 is a multivariate gain
function of ${\bf u}\in \left\{ -1,1\right\} ^{n}$.

There is a similar result also for the variance in (1), based on
Cauchy-Schwarz inequality.\bigskip

{\bf Lemma 2}: $s\geq ({\bf y}-\overline{y}{\bf 1}_{n})^{\prime }{\bf u/}%
\sqrt{n}\ \ $\ for$\ \ {\bf u}^{\prime }{\bf u=}1{\bf .\bigskip }$

We note that Lemmas 1 and 2 represent particular cases of H\"{o}lder
inequality, see Choulakian ( (2016).\bigskip

{\bf Definition 1}: We define the relative contribution of an element $y_{i}$
to $d,$ $LAD$ and $s^{2}$, respectively, to be%
\[
RC_{d}(y_{i})=\frac{|y_{i}-\overline{y}|}{n\ d},
\]

\[
RC_{s^{2}}(y_{i})=\frac{|y_{i}-\overline{y}|^{2}}{n\ s^{2}},
\]

\[
RC_{LAD}(y_{i})=\frac{|y_{i}-median|}{n\ AD}.
\]%
Then the following inequalities are true%
\[
0\leq RC_{d}(y_{i})\leq 0.5,
\]%
\[
0\leq RC_{s^{2}}(y_{i})<1,
\]%
\[
0\leq RC_{LAD}(y_{i})\leq 1;
\]%
from which we conclude that the most robust dispersion measure among the
three dispersion measures, based on the relative contribution criterion, is $%
d.$

We note that the inequality, $0\leq RC_{s^{2}}(y_{i})<1,$ is a weaker
variant of Laguerre-Samuelson inequality; see for instance, Jensen (1999),
whose MS thesis presents nine different proofs.

We have\bigskip

{\bf Definition 2}: An element $x_{i}=y_{i}-\overline{y}$ is a heavyweight
if $RC_{d}(y_{i})=0.5;$ that is, $|x_{i}|=|y_{i}-\overline{y}|=nd/2.\bigskip
$

We note that a heavyweight element attains the upper bound of $RC_{d}(y_{i}),
$ but it never attains the upper bound of $RC_{s^{2}}(y_{i})$ and $%
RC_{LAD}(y_{i}).$

\section{2-way interactions of a correspondence matrix}

Let ${\bf P=(}p_{ij})\ $be a correspondence matrix; that is, $p_{ij}\geq 0$
for $i\in I$\ and $j\in J=\left\{ 1,2,...,m\right\} $ and $%
\sum_{j=1}^{m}\sum_{i=1}^{n}p_{ij}=1.$ As usual, we define $p_{i\ast
}=\sum_{j=1}^{m}p_{ij}$ and $p_{\ast j}=\sum_{i=1}^{n}p_{ij}$. Let ${\bf P}%
_{1}{\bf =(}x_{ij}=p_{ij}-p_{i\ast }p_{\ast j})$ for $i\in I$\ and $j\in J;$
then ${\bf P}_{1}$ represents the residual matrix of ${\bf P}$ with respect
to the independence model $(p_{i\ast }p_{\ast j})$. In the jargon of
statistics, the cell $x_{ij}$ represents the multiplicative 2-way
interaction of the cell ($i,j)\in I\times J.$\ ${\bf P}_{1}$ is
double-centered%
\begin{equation}
{\bf P}_{1}{\bf 1}_{m}={\bf 0}_{n}\text{\ \ \ \ \ and\ \ \ \ \ }{\bf P}_{1}%
{\bf ^{\prime }1}_{n}={\bf 0}_{n}.
\end{equation}%
From (5) we get%
\begin{eqnarray}
\sum_{i\in S}x_{ij} &=&-\sum_{i\in \overline{S}}x_{ij}\text{\ \ \ \ for }%
j\in J, \\
\sum_{j\in T}x_{ij} &=&-\sum_{j\in \overline{T}}x_{ij}\text{\ \ \ \ for }%
i\in I,
\end{eqnarray}%
for $T\subset J.$ From (6) and (7), we get%
\begin{eqnarray}
\sum_{j\in T}\sum_{i\in S}x_{ij} &=&-\sum_{j\in T}\sum_{i\in \overline{S}%
}x_{ij}, \\
&=&-\sum_{i\in S}\sum_{j\in \overline{T}}x_{ij} \\
&=&\sum_{i\in \overline{S}}\sum_{j\in \overline{T}}x_{ij}.
\end{eqnarray}%
We define the cut-norm of ${\bf P}_{1}$ to be
\[
||{\bf P}_{1}{\bf ||}_{\boxdot }=\max_{S,T}\sum_{j\in T}\sum_{i\in
S}x_{ij}=\sum_{j\in T_{opt}}\sum_{i\in S_{opt}}x_{ij}.
\]%
The cut-norm $||{\bf P}_{1}{\bf ||}_{\boxdot }$ is a well known quantity in
theoretical computer science, because of its relationship to the famous
Grothendieck inequality, which is based on $||{\bf P}_{1}||_{\infty
\rightarrow 1},$ see among others Khot and Naor (2012).

The matrix ${\bf P}_{1}$ can be considered as the starting point in taxicab
correspondence analysis, an L$_{1}$ variant of correspondence analysis , see
Choulakian (2006). The optimization criterion in TCA of ${\bf P}$ or ${\bf P}%
_{1}$\ is based on taxicab matrix norm, which is a combinatorial
optimization problem

\begin{eqnarray}
\delta _{1} &=&||{\bf P}_{1}||_{\infty \rightarrow 1}=||{\bf P}_{1}^{\prime
}||_{\infty \rightarrow 1}  \nonumber \\
&=&\max_{{\bf u\in
\mathbb{R}
}^{m}}\frac{\left\vert \left\vert {\bf P}_{1}{\bf u}\right\vert \right\vert
_{1}}{\left\vert \left\vert {\bf u}\right\vert \right\vert _{\infty }}=\max_{%
{\bf v\in
\mathbb{R}
}^{n}}\ \frac{\left\vert \left\vert {\bf P}_{1}{\bf ^{\prime }v}\right\vert
\right\vert _{1}}{\left\vert \left\vert {\bf v}\right\vert \right\vert
_{\infty }}=\max_{{\bf u\in
\mathbb{R}
}^{m},{\bf v\in
\mathbb{R}
}^{n}}\frac{{\bf v}^{\prime }{\bf P}_{1}{\bf u}}{\left\vert \left\vert {\bf u%
}\right\vert \right\vert _{\infty }\left\vert \left\vert {\bf v}\right\vert
\right\vert _{\infty }},  \nonumber \\
&=&\max ||{\bf P}_{1}{\bf u||}_{1}\ \ \text{subject to }{\bf u}\in \left\{
-1,+1\right\} ^{m},  \nonumber \\
&=&\max ||{\bf P}_{1}{\bf ^{\prime }v||}_{1}\ \ \text{subject to }{\bf v}\in
\left\{ -1,+1\right\} ^{n},  \nonumber \\
&=&\max {\bf v}^{\prime }{\bf P}_{1}{\bf u}\text{ \ subject to \ }{\bf u}\in
\left\{ -1,+1\right\} ^{m},{\bf v}\in \left\{ -1,+1\right\} ^{n}, \\
&=&{\bf v}_{1}^{\prime }{\bf P}_{1}{\bf u}_{1}.
\end{eqnarray}%
In data analysis, the vectors ${\bf v}_{1}${\bf \ }and{\bf \ }${\bf u}_{1}$
are interpreted as taxicab principal axes and $\delta _{1}$ as first taxicab
dispersion. So we can compute the projection of the rows (resp. the columns)
of ${\bf P}_{1}$ on the taxicab principal axis ${\bf u}_{1}$ (resp. ${\bf v}%
_{1})$ to be
\begin{eqnarray}
{\bf a}_{1} &=&{\bf P}_{1}{\bf u}_{1} \\
{\bf b}_{1} &=&{\bf P}_{1}^{\prime }{\bf v}_{1}.
\end{eqnarray}%
Equation (12) implies%
\begin{eqnarray}
{\bf v}_{1} &=&sign({\bf a}_{1}), \\
{\bf u}_{1} &=&sign({\bf b}_{1}),
\end{eqnarray}%
named transition formulas, see Choulakian (2006, 2016). We also note the
following identities%
\begin{eqnarray}
{\bf 1}_{n}^{\prime }{\bf a}_{1} &=&0\text{ \ \ \ and \ \ }\delta _{1}=||%
{\bf a}_{1}||_{1}, \\
{\bf 1}_{m}^{\prime }{\bf b}_{1} &=&0\text{ \ \ \ and \ \ }\delta _{1}=||%
{\bf b}_{1}||_{1}.
\end{eqnarray}

Using the above results, we get the following \bigskip

{\bf Lemma 3}: (4-equal parts property) The norm $\delta _{1}=||{\bf P}%
_{1}||_{\infty \rightarrow 1}=4||{\bf P}_{1}{\bf ||}_{\boxdot }\geq
4\sum_{j\in T}\sum_{i\in S}x_{ij}.\bigskip $

In data analysis, Lemma 3 implies balanced 2-blocks seriation of ${\bf P}%
_{1};$ see example 1. The subsets $T_{opt}$ and $S_{opt}$ are positively
associated and $||{\bf P}_{1}{\bf ||}_{\boxdot }=\sum_{j\in
T_{opt}}\sum_{i\in S_{opt}}x_{ij}$; similarly the subsets $\overline{T}_{opt}
$ and $\overline{S}_{opt}$ are positively associated and $||{\bf P}_{1}{\bf %
||}_{\boxdot }=\sum_{j\in \overline{T}_{opt}}\sum_{i\in \overline{S}%
_{opt}}x_{ij}.$ While the subsets $\overline{T}_{opt}$ and $S_{opt}$ are
negatively associated and $||{\bf P}_{1}{\bf ||}_{\boxdot }=-\sum_{j\in
\overline{T}_{opt}}\sum_{i\in S_{opt}}x_{ij};$ similarly the subsets $T_{opt}
$ and $\overline{S}_{opt}$ are negatively associated and $||{\bf P}_{1}{\bf %
||}_{\boxdot }=-\sum_{j\in T_{opt}}\sum_{i\in \overline{S}_{opt}}x_{ij}.$
Liiv (2010) presents an interesting overview of seriation.

Using Definition 2, we get\bigskip

{\bf Definition 3}: The relative contribution of the row $i$ to $\delta _{1}$
(respectively of the column $j$ to $\delta _{1}$) is%
\[
RC_{_{1}}(\func{row}\ i)=\frac{|{\bf a}_{1}(i)|}{\delta _{1}}\text{ and }%
RC_{_{1}}(\func{col}\ j)=\frac{|{\bf b}_{1}(i)|}{\delta _{1}}.
\]%
\bigskip

We have%
\[
0\leq RC_{1}(\func{row}\ i)\text{ and }RC_{1}(\func{col}\ j)\leq 0.5.
\]%
\bigskip

{\bf Definition 4}: a) On the first taxicab principlal axis the row $i$ is
heavyweight if $RC_{1}(\func{row}\ i)=0.5,$ and the column $j$ is
heavyweight if $RC_{1}(\func{col}\ j)=0.5.$

b) On the first taxicab principlal axis the cell $(i,j)$ is heavyweight if
and only if both row $i$ and column $j$ are heavyweights; and in this case $%
RC_{1}(p_{ij}-p_{i\ast }p_{\ast j})=\frac{|p_{ij}-p_{i\ast }p_{\ast j}|}{%
\delta _{1}}=0.25.\bigskip $

For an application of Definitions 3 and 4 see Choulakian (2008a).

Using Wedderburn's rank-1 reduction rule, see Choulakian (2016), we
construct the 2nd residual matrix ${\bf P}_{2}={\bf (}x_{ij}=p_{ij}-p_{i\ast
}p_{\ast j}-\frac{a_{1}(i)b_{1}(j)}{\sigma _{1}}),$ and repeat the above
procedure. After $k=rank({\bf P}_{1})$ iterations, we decompose the
correspondence matrix ${\bf P}$ into $(k+1)$ bilinear parts%
\[
p_{ij}=p_{i\ast }p_{\ast j}+\sum_{\alpha =1}^{k}\frac{a_{\alpha
}(i)b_{\alpha }(j)}{\delta _{\alpha }},
\]%
named taxicab singular value decomposition; which can be rewritten, similar
to data reconstruction formula in correspondence analysis (CA), as

\[
p_{ij}=p_{i\ast }p_{\ast j}(1+\sum_{\alpha =1}^{k}\frac{f_{\alpha
}(i)g_{\alpha }(j)}{\delta _{\alpha }}),
\]%
where $f_{\alpha }(i)=a_{\alpha }(i)/p_{i\ast }$ and $g_{\alpha
}(j)=b_{\alpha }(j)/p_{\ast j}.$ CA and TCA satisfy an important invariance
property: columns (or rows) with identical profiles (conditional
probabilities) receive identical factor scores $g_{\alpha }(j)$ (or $%
f_{\alpha }(i))$. The factor scores are used in the graphical displays.
Moreover, merging of identical profiles does not change the results of the
data analysis: This is named the {\it principle of equivalent partitioning}
by Nishisato (1984); it includes the famous invariance property named {\it %
principle of distributional equivalence}, on which Benz\'{e}cri (1973)
developed CA.

In the next subsections we shall present two examples, where taxicab
correspondence analysis (TCA) is applied. The first data set is a small
contingency table taken from Beh and Lombardo (2014), for which we present
the details of the computation explaing the contents of section 3. The
second data set is a networks affinity matrix from Faust (2005). For both
data sets we compare CA and TCA maps.

The theory of CA can be found, among others, in Benz\'{e}cri (1973, 1992),
Greenacre (1984), Gifi (1990), Le Roux and Rouanet (2004), Murtagh (2005),
and Nishisato (2007); the recent book, authored by Beh and Lombardi (2014),
presents a panoramic review of CA and related methods.

\subsection{Selikoff's asbestos data set}

Table 1, taken from Beh and Lombardo (2014), is a contingency table {\bf Y}
of size $5\times 4$ cross-classifying 1117 New York workers with
occupational exposure to asbestos; the workers are classified according to
the number of exposure in years (five categories) and the asbestos grade
diagnosed (four categories). Figures 1 and 2 display the maps obtained by CA
and TCA: almost no difference between them. Here, we present the details of
the computation for TCA. Table 2 presents the residual correspondence table $%
{\bf P}_{1}$ with respect to the independence model, where we see diagonal
2-blocks seriation of ${\bf P}_{1}$ with$:$ $S_{opt}=\left\{ \text{row1, row2%
}\right\} $ is positively associated with $T_{opt}=\left\{ \text{column1}%
\right\} $ and the cut-norm $||{\bf P}_{1}{\bf ||}_{\boxdot
}=(0.1181+0.0151)=0.1332;$ similarly, $\overline{S}_{opt}=\left\{ \text{%
row3, row4, row5}\right\} $ is positively associated with \\ $\overline{T}%
_{opt}=\left\{ \text{column2, column3, column4}\right\} $ and $||{\bf P}_{1}%
{\bf ||}_{\boxdot }=\sum_{j\in \overline{T}_{opt}}\sum_{i\in \overline{S}%
_{opt}}x_{ij}=0.0087+...+0.0202=0.1332.\ $Note that the elements in the
positively associated diagonal blocks have in majority positive values;
while the elements in the negatively associated diagonal blocks have in
majority negative values. The last three columns and the last three rows of
Table 2 display principal axes (${\bf v}_{1}${\bf \ }and{\bf \ }${\bf u}_{1})
$, coordinates of the projected points (${\bf a}_{1}${\bf \ }and{\bf \ }$%
{\bf b}_{1})$ and coordinates of TCA factor scores (${\bf f}_{1}${\bf \ }and%
{\bf \ }${\bf g}_{1}).$

Table 3 shows the 2nd residual correspondence matrix ${\bf P}_{2}$, where we
note that its first column is zero, because column 1 is heavyweight in ${\bf %
P}_{1}:$ the $RC_{1}^{TCA}(G0)=0.5$, see Choulakian (2008a). We see that
columns (3 and 4) are positively associated with rows (1 and 5); similarly
column 2 is positively associated with rows 2 to 4. It is difficult to
interpret the diagonal balanced 2-blocks seriation in Table 3; however, the
map in Figure 2 is interpretable, it shows a Guttman effect known as
horseshoe or parabola.

\begin{tabular}{|l||l|l|l|l||l||l||}
\multicolumn{7}{l}{{\bf Table 1: Selikoff's Asbestos contingency table of
size }$5\times 4${\bf .}} \\ \hline
Exposure & \multicolumn{6}{||l||}{Asbestos Grade Diagnosed} \\ \cline{2-7}
in years & \multicolumn{1}{||l|}{None (G0)} & Grade 1(G1) & Grade 2 (G2) &
Grade 3 (G3) & Total & $p_{i\ast }$ \\ \hline
0-9 & \multicolumn{1}{||l|}{310} & 36 & 0 & 0 & 346 & 0.3098 \\ \hline
10-19 & \multicolumn{1}{||l|}{212} & 158 & 9 & 0 & 379 & 0.3393 \\ \hline
20-29 & \multicolumn{1}{||l|}{21} & 35 & 17 & 4 & 77 & 0.0689 \\ \hline
30-39 & \multicolumn{1}{||l|}{25} & 102 & 49 & 18 & 194 & 0.1737 \\ \hline
40+ & \multicolumn{1}{||l|}{7} & 35 & 51 & 28 & 121 & 0.1083 \\ \hline\hline
Total & 575 & 366 & 126 & 50 & 1117 & 1 \\ \hline\hline
$p_{\ast j}$ & 0.5148 & 0.3277 & 0.1128 & 0.0448 & 1 &  \\ \hline\hline
\end{tabular}

\bigskip

{\small \begin{tabular}{|l||ll|l|l||l||l|l|l|}
\multicolumn{9}{l}{{\bf Table 2: Balanced 2-blocks seriation of }${\bf P}_{1}%
{\bf =(}p_{ij}-p_{i\ast }p_{\ast j})${\bf \ of size }$5\times 4${\bf .}} \\
\hline
& \multicolumn{8}{||l|}{Asbestos Grade Diagnosed} \\ \hline
Exposure & None & Grade 1 & Grade 2 & Grade 3 & Total &
\multicolumn{1}{||l|}{${\bf v}_{1}$} & ${\bf a}_{1}$ & ${\bf f}_{1}$ \\
\cline{6-9}
in years & (G0) & \multicolumn{1}{|l|}{(G1)} & (G2) & (G3) &  &
\multicolumn{1}{||l|}{} &  &  \\ \hline
0-9 & \multicolumn{1}{||l|}{${\bf 0.1181}$} & \multicolumn{1}{|l|}{$-0.0693$}
& $-0.0349$ & $-0.0139$ & $0$ & \multicolumn{1}{||l|}{$-1$} & $-0.2362$ & $%
-0.7624$ \\ \hline
10-19 & \multicolumn{1}{||l|}{${\bf 0.0151}$} & \multicolumn{1}{|l|}{$0.0303$%
} & $-0.0302$ & $-0.0152$ & $0$ & \multicolumn{1}{||l|}{$-1$} & $-0.0303$ & $%
-0.0892$ \\ \hline
20-29 & \multicolumn{1}{||l|}{$-0.0167$} & \multicolumn{1}{|l|}{${\bf 0.0087}
$} & ${\bf 0.0074}$ & ${\bf 0.0005}$ & $0$ & \multicolumn{1}{||l|}{$1$} & $%
0.0334$ & $0.4841$ \\ \hline
30-39 & \multicolumn{1}{||l|}{$-0.0670$} & \multicolumn{1}{|l|}{${\bf 0.0344}
$} & ${\bf 0.0243}$ & ${\bf 0.0083}$ & $0$ & \multicolumn{1}{||l|}{$1$} & $%
0.1340$ & $0.7718$ \\ \hline
40+ & \multicolumn{1}{||l|}{$-0.0495$} & \multicolumn{1}{|l|}{${\bf -0.0042}$%
} & ${\bf 0.0334}$ & ${\bf 0.0202}$ & $0$ & \multicolumn{1}{||l|}{$1$} & $%
0.0990$ & $0.9138$ \\ \hline\hline
Total & $0$ & \multicolumn{1}{|l|}{$0$} & $0$ & $0$ & $0$ &
\multicolumn{1}{||l|}{} & \multicolumn{2}{|l|}{$\delta _{1}=4\times 0.1332$}
\\ \hline
${\bf u}_{1}$ & $-1$ & \multicolumn{1}{|l|}{$1$} & $1$ & $1$ &  &  &
\multicolumn{2}{|l|}{$||{\bf P}_{1}{\bf ||}_{\boxdot }=0.1332$} \\
\hline\hline
${\bf b}_{1}$ & $-0.2664$ & \multicolumn{1}{|l|}{$0.0780$} & $0.1302$ & $%
0.0582$ &  &  & $0.1332$ & \multicolumn{1}{||l|}{$|-0.1332|$} \\ \hline\hline
${\bf g}_{1}$ & $-0.5175$ & \multicolumn{1}{|l|}{$0.2380$} & $1.1553$ & $%
1.2981$ &  &  & $|-0.1332|$ & \multicolumn{1}{||l|}{$0.1332$} \\ \hline\hline
\end{tabular}%
}

\begin{tabular}{|l||l|l|l|l||l||l|l|l|}
\multicolumn{9}{l}{{\bf Table 3: Balanced 2-blocks seriation of }${\bf P}%
_{2}={\bf (}p_{ij}-p_{i\ast }p_{\ast j}-\frac{a_{1i}b_{1j}}{\delta _{1}})$%
{\bf \ of size }$5\times 4${\bf .}} \\ \hline
\multicolumn{1}{|l||}{} & \multicolumn{8}{||l|}{Asbestos Grade Diagnosed} \\
\cline{2-9}
Exposure & None & Grade 1 & Grade 2 & Grade 3 & Total &
\multicolumn{1}{||l|}{${\bf v}_{2}$} & ${\bf a}_{2}$ & ${\bf f}_{2}$ \\
\cline{6-9}
\multicolumn{1}{|l|}{in years} & (G0) & (G1) & (G2) & (G3) &  &
\multicolumn{1}{||l|}{} &  &  \\ \hline
\multicolumn{1}{|l|}{0-9} & \multicolumn{1}{||l|}{$0$} & $-0.0347$ & ${\bf %
0.0228}$ & ${\bf 0.0119}$ & $0$ & \multicolumn{1}{||l|}{$1$} & $0.0694$ & $%
0.2241$ \\ \hline
\multicolumn{1}{|l|}{40+} & \multicolumn{1}{||l|}{$0$} & $-0.0186$ & ${\bf %
0.0092}$ & ${\bf 0.0094}$ & $0$ & \multicolumn{1}{||l|}{$1$} & $0.0372$ & $%
0.3443$ \\ \hline
\multicolumn{1}{|l|}{10-19} & \multicolumn{1}{||l|}{$0$} & ${\bf 0.0347}$ & $%
-0.0228$ & $-0.0119$ & $0$ & \multicolumn{1}{||l|}{$-1$} & $-0.0694$ & $%
-0.2046$ \\ \hline
\multicolumn{1}{|l|}{20-29} & \multicolumn{1}{||l|}{$0$} & ${\bf 0.0038}$ & $%
-0.0007$ & $-0.0031$ & $0$ & \multicolumn{1}{||l|}{$-1$} & $-0.0076$ & $%
-0.1121$ \\ \hline
\multicolumn{1}{|l|}{30-39} & \multicolumn{1}{||l|}{$0$} & ${\bf 0.0148}$ & $%
-0.0085$ & $-0.0063$ & $0$ & \multicolumn{1}{||l|}{$-1$} & $-0.0296$ & $%
-0.1703$ \\ \hline\hline
\multicolumn{1}{|l||}{Total} & $0$ & $0$ & $0$ & $0$ & $0$ &
\multicolumn{1}{||l|}{} & \multicolumn{2}{|l|}{$\delta _{2}=4\times 0.0533$}
\\ \hline
\multicolumn{1}{|l||}{${\bf u}_{2}$} & $\pm 1$ & $-1$ & $1$ & $1$ &  &  &
\multicolumn{2}{|l|}{$||{\bf P}_{2}{\bf ||}_{\boxdot }=0.0533$} \\
\hline\hline
\multicolumn{1}{|l||}{${\bf b}_{2}$} & $0$ & $-0.1066$ & $0.0640$ & $0.0426$
&  &  & $|-0.0533|$ & \multicolumn{1}{||l|}{$0.0533$} \\ \hline\hline
\multicolumn{1}{|l||}{${\bf g}_{2}$} & $0$ & $-0.3257$ & $0.5681$ & $0.9521$
&  &  & $0.0533$ & \multicolumn{1}{||l|}{$|-0.0533|$} \\ \hline\hline
\end{tabular}%
\bigskip \bigskip

\bigskip


\begin{figure}
  \centering
  \includegraphics[width=3.8035in]{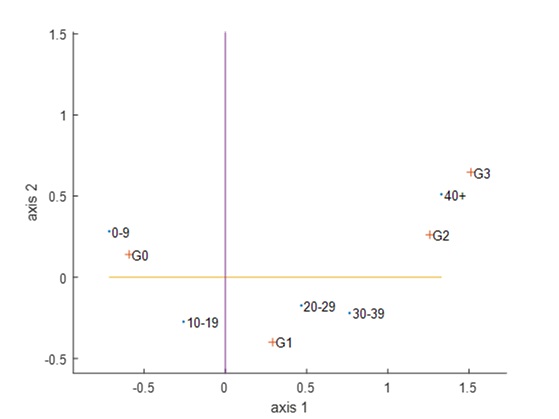}\\
  \caption{CA map of asbestos exposure data.}
  \end{figure}


\begin{figure}
  \centering
  \includegraphics[width=3.8035in]{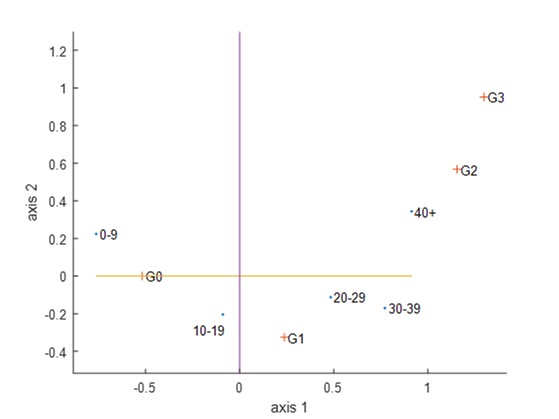}\\
  \caption{TCA map of asbestos exposure data.}
\end{figure}

\bigskip

\subsection{Western Hemisphere countries and their memberships in trade and
treaty organizations}

Table 4 presents a two-mode affiliation network matrix ${\bf Z}=(z_{ij})$ of
size $22\times 15$ taken from Faust (2005). The 22 rows represent 22
countries and the 15 columns the regional trade and treaty organizations,
described in Appendix A. The country $i$ is a member of the organization $j$
if $z_{ij}=1$; and $z_{ij}=0$ means the country $i$ is not a member of the
organization $j$. Faust (2005) visualized this data by correspondence
analysis, see Figure 3, which is quite cluttered. She interpreted the first
two principal dimensions by examining the factor scores of the countries and
summarized the results in 3 points:

a) The first dimension contrasts South American countries and organizations
on the one hand, and Central American countries and organizations on the
other hand.

b) The second dimension clearly distinguishes Canada and the United States
(both North American countries) along with NAFTA from other countries and
organizations. In CA, the relative contribution of Canada (resp. US) to the
second axis is $RC_{2}^{CA}(Canada)=RC_{2}^{CA}(US)=0.409$, and $%
RC_{2}^{CA}(NAFTA)=0.821.$

c) Organizations (SELA, OAS, and IDB) are in the center because they have
membership profiles that are similar to the marginal profile: almost all
countries belong to (SELA, OAS, and IDB), see Table 4.

Figure 4 provides the TCA map, which is much more interpretable than the
corresponding CA map in Figure 3; where we see that, additionally to the
three points mentioned by Faust (2005), the south american countries are
divided into two groups, northern (Venezuela, Bolivia, Peru and Ecuador) and
southern countries (Brazil, Uruguay, Argentina, Paraguay and Chile).
Furthermore, the contributions of the points Canada, the United States, and
NAFTA to the second axis are not substantial compared to CA: $%
RC_{2}^{TCA}(Canada)=RC_{2}^{TCA}(US)=0.088$, and $RC_{2}^{TCA}(NAFTA)=0.10.$
This shows the robustness of TCA due to the robustness of the $\delta $
statistic.

It is well known that, CA is very sensitive to some particularities of a
data set; further, how to identify and handle these is an open unresolved
problem. However, for contingency tables Choulakian (2017) enumerated three
under the umbrella of sparse contingency tables: rare observations,
zero-block structure and relatively high-valued cells. It is evident that
this data set has specially three rare observations (NAFTA, CANADA and USA),
which determine the 2nd dimension of CA. A row or a column is considered
rare, if its marginal probablity is quite small.

\begin{figure}
  \centering
  \includegraphics[width=3.8035in]{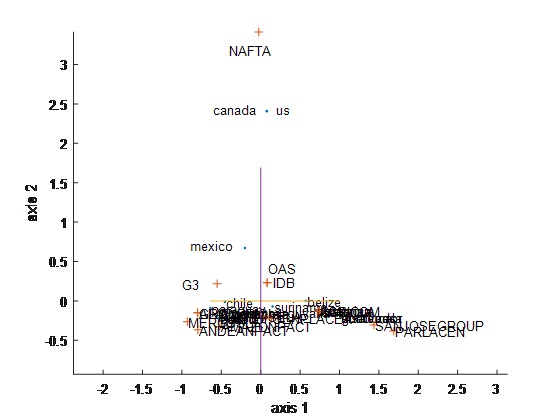}\\
  \caption{CA map of Western Hemisphere affinity network.}
\end{figure}
\begin{figure}
  \centering
  \includegraphics[width=3.8035in]{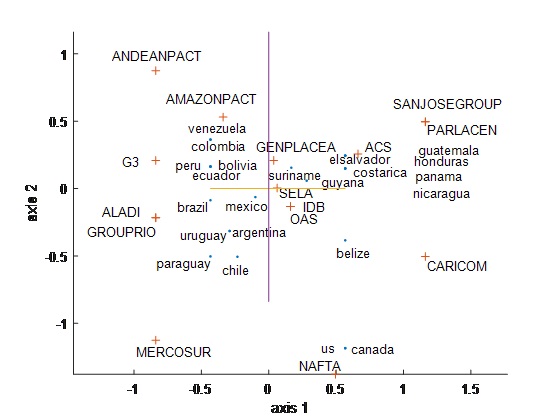}\\
  \caption{TCA map of Western Hemisphere affinity network.}
\end{figure}


\begin{tabular}{l|lllllllllllllll|l}
\multicolumn{17}{l}{\bf Table 4: Sociomatrix of American countries and their
memberships.} \\ \hline
Countries & \multicolumn{16}{|l}{Regional Trade and Treaty Organizations} \\
& 1 & 2 & 3 & 4 & 5 & 6 & 7 & 8 & 9 & 10 & 11 & 12 & 13 & 14 & 15 & Sum \\
\hline
Argentina & 0 & 1 & 0 & 0 & 0 & 1 & 1 & 0 & 1 & 1 & 0 & 1 & 0 & 0 & 1 & 7 \\
Belize & 1 & 0 & 0 & 0 & 1 & 0 & 0 & 0 & 1 & 0 & 0 & 1 & 0 & 0 & 1 & 5 \\
Bolivia & 0 & 1 & 1 & 1 & 0 & 1 & 1 & 0 & 1 & 0 & 0 & 1 & 0 & 0 & 1 & 8 \\
Brazil & 0 & 1 & 1 & 0 & 0 & 1 & 1 & 0 & 1 & 1 & 0 & 1 & 0 & 0 & 1 & 8 \\
Canada & 0 & 0 & 0 & 0 & 0 & 0 & 0 & 0 & 1 & 0 & 1 & 1 & 0 & 0 & 0 & 3 \\
Chile & 0 & 1 & 0 & 0 & 0 & 0 & 1 & 0 & 1 & 0 & 0 & 1 & 0 & 0 & 1 & 5 \\
Colombia & 1 & 1 & 1 & 1 & 0 & 1 & 1 & 1 & 1 & 0 & 0 & 1 & 0 & 0 & 1 & 10 \\
CostaRica & 1 & 0 & 0 & 0 & 0 & 1 & 0 & 0 & 1 & 0 & 0 & 1 & 0 & 1 & 1 & 6 \\
Ecuador & 0 & 1 & 1 & 1 & 0 & 1 & 1 & 0 & 1 & 0 & 0 & 1 & 0 & 0 & 1 & 8 \\
ElSalvador & 1 & 0 & 0 & 0 & 0 & 1 & 0 & 0 & 1 & 0 & 0 & 1 & 1 & 1 & 1 & 7
\\
Guatemala & 1 & 0 & 0 & 0 & 0 & 1 & 0 & 0 & 1 & 0 & 0 & 1 & 1 & 1 & 1 & 7 \\
Guyana & 1 & 0 & 1 & 0 & 1 & 1 & 0 & 0 & 1 & 0 & 0 & 1 & 0 & 0 & 1 & 7 \\
Honduras & 1 & 0 & 0 & 0 & 0 & 1 & 0 & 0 & 1 & 0 & 0 & 1 & 1 & 1 & 1 & 7 \\
Mexica & 1 & 1 & 0 & 0 & 0 & 1 & 1 & 1 & 1 & 0 & 1 & 1 & 0 & 0 & 1 & 9 \\
Nicaragua & 1 & 0 & 0 & 0 & 0 & 1 & 0 & 0 & 1 & 0 & 0 & 1 & 0 & 1 & 1 & 6 \\
Panama & 1 & 0 & 0 & 0 & 0 & 1 & 0 & 0 & 1 & 0 & 0 & 1 & 0 & 1 & 1 & 6 \\
Paraguay & 0 & 1 & 0 & 0 & 0 & 0 & 1 & 0 & 1 & 1 & 0 & 1 & 0 & 0 & 1 & 6 \\
Peru & 0 & 1 & 1 & 1 & 0 & 1 & 1 & 0 & 1 & 0 & 0 & 1 & 0 & 0 & 1 & 8 \\
Suriname & 1 & 0 & 1 & 0 & 0 & 0 & 0 & 0 & 1 & 0 & 0 & 1 & 0 & 0 & 1 & 5 \\
UnitedStates & 0 & 0 & 0 & 0 & 0 & 0 & 0 & 0 & 1 & 0 & 1 & 1 & 0 & 0 & 0 & 3
\\
Uruguay & 0 & 1 & 0 & 0 & 0 & 1 & 1 & 0 & 1 & 1 & 0 & 1 & 0 & 0 & 1 & 7 \\
Venezuela & 1 & 1 & 1 & 1 & 0 & 1 & 1 & 1 & 1 & 0 & 0 & 1 & 0 & 0 & 1 & 10
\\ \hline
\multicolumn{1}{|l|}{Sum} & 12 & \multicolumn{1}{|l}{11} &
\multicolumn{1}{|l}{8} & \multicolumn{1}{|l}{5} & \multicolumn{1}{|l}{2} &
\multicolumn{1}{|l}{16} & \multicolumn{1}{|l}{11} & \multicolumn{1}{|l}{3} &
\multicolumn{1}{|l}{22} & \multicolumn{1}{|l}{5} & \multicolumn{1}{|l}{3} &
\multicolumn{1}{|l}{22} & \multicolumn{1}{|l}{3} & \multicolumn{1}{|l}{6} &
\multicolumn{1}{|l|}{20} & \multicolumn{1}{|l|}{149} \\ \hline
\end{tabular}%
\bigskip

\subsection{Maximal interaction two-mode clustering of continuous data}

Schepers, Bock and Van Mechelen (2017) discussed maximum interaction
two-mode clustering of continuous data. By generalizing their objective
function, we want to show that the results of this section can be considered
a particular robust L$_{1}\ $variant of their approach. Let ${\bf Y=(}%
y_{ij})\ $be a 2-way array for $i\in I,$\ $j\in J$. As usual, we define, for
instance, $\overline{y}_{\ast j}=\sum_{i=1}^{n}\frac{y_{ij}}{n}$ and $%
\overline{y}_{\ast \ast }=\sum_{j=1}^{m}\sum_{i=1}^{n}\frac{y_{ij}}{mn}.$
Let ${\bf X=(}x_{ij})$ be the additive double-centered array, where
\[
x_{ij}=y_{ij}-\overline{y}_{i\ast }-\overline{y}_{\ast j}+\overline{y}_{\ast
\ast }.
\]%
In the jargon of statistics, the cell $x_{ij}$ represents the additive 2-way
interaction of the cell ($i,j)\in I\times J.$ The matrix ${\bf X}$ is
double-centered, and it satisfies equations (6) through (10). Let $%
I=U_{\alpha =1}^{r}S_{\alpha }$ be an $r$-partition of $I$ and $J=U_{\beta
=1}^{c}T_{\beta }$ be a $c$-partition of $J.$ We consider the following
maximization of the overall interaction problem for $p\geq 1$%
\[
f_{p}(S_{\alpha },T_{\beta }:\alpha =1,...,r\ \text{and}\ \beta
=1,...,c)=\sum_{\alpha =1}^{r}\sum_{\beta =1}^{c}|S_{\alpha }|\ |T_{\beta
}|\ g_{p}(\alpha ,\beta ),
\]%
where $|S_{\alpha }|$ is the cardinality of the set $S_{\alpha }$ and%
\[
g_{p}(\alpha ,\beta )=\left( |\frac{\sum_{i\in S_{\alpha }}\sum_{j\in
T_{\beta }}x_{ij}}{|S_{\alpha }|\ |T_{\beta }|\ }|\right) ^{p}.
\]%
When $p=2$, then maximizing $f_{2}(S_{\alpha },T_{\beta }:\alpha =1,...,r\ $%
and$\ \beta =1,...,c),$ named maximal overall interaction, is the criterion
computed in Schepers et. al (2017). When $p=1,\ r=c=2,$ then maximizing $%
f_{1}(S_{\alpha },T_{\beta }:\alpha =1,...,2\ $and$\ \beta =1,...,2)=||{\bf %
X||}_{\infty \rightarrow 1}=4||{\bf X||}_{\boxdot }$ by Lemma 3, which is
the criterion computed in TCA.

\section{Triple-centered arrays}

To motivate our subject, we start with an example. Let ${\bf Y=(}y_{ijk})\ $%
be a 3-way array for $i\in I,$\ $j\in J$ and $k\in K=\left\{
1,2,...,t\right\} .$ As usual, we define, for instance, $\overline{y}%
_{ij\ast }=\sum_{k=1}^{t}y_{ijk}/t,,$ $\overline{y}_{\ast j\ast
}=\sum_{k=1}^{t}\sum_{i=1}^{n}\frac{y_{ijk}}{tn}$ and $\overline{y}_{\ast
\ast \ast }=\sum_{k=1}^{t}\sum_{j=1}^{m}\sum_{i=1}^{n}\frac{y_{ijk}}{tmn}.$
Let ${\bf X=(}x_{ijk})$ be the triple-centered array, where
\[
x_{ijk}=y_{ijk}-\overline{y}_{ij\ast }-\overline{y}_{i\ast k}-\overline{y}%
_{\ast jk}+\overline{y}_{i\ast \ast }+\overline{y}_{\ast j\ast }+\overline{y}%
_{\ast \ast k}-\overline{y}_{\ast \ast \ast }.
\]%
In the jargon of statistics, the cell $x_{ijk}$ represents the additive
3-way interaction of the cell ($i,j,k)\in I\times J\times K.$ The tensor $%
{\bf X}$ is triple-centered; that is,%
\[
\sum_{k=1}^{t}x_{ijk}=\sum_{j=1}^{n}x_{ijk}=\sum_{i=1}^{m}x_{ijk}=0.
\]%
A generalization of Lemma 3 is\bigskip

{\bf Lemma 4}: (8-equal parts property) The tensor norm%
\begin{eqnarray*}
||{\bf X}||_{(\infty ,\infty )\rightarrow 1} &=&\max \sum_{k\in K}\sum_{j\in
J}\sum_{i\in I}w_{k}v_{j}u_{i}x_{ijk}\ \text{subject to}\ {\bf u\times v}%
\times {\bf w}\in \left\{ -1,+1\right\} ^{m\times n\times t} \\
&=&8\sum_{k\in W_{opt}}\sum_{j\in T_{opt}}\sum_{i\in S_{opt}}x_{ijk}\geq
8\sum_{k\in W}\sum_{j\in T}\sum_{i\in S}x_{ijk},
\end{eqnarray*}
where $W\subset K.$

The proof is similar to the proof of Lemma 3.

Lemma 4 can easily be generalized to higher-way arrays.

\subsection{Conclusion}

This essay is an attempt to emphasize the following two points.

First, we showed the optimality and robustness of the mean absolute
deviations about the mean, its interpretation, and its generalization to
higher-way arrays. A key notion in describing its robustness is that the
relative contribution of a point is bounded by 50\%.

Second, within the framework of TCA, we showed that the following three
identities $\delta _{1}=||{\bf P}_{1}||_{\infty \rightarrow 1}=4||{\bf P}_{1}%
{\bf ||}_{\boxdot }$ reveal three different but related aspects of TCA: a)  $%
\delta _{1}$, computed in (17) and (18), represents the mean of absolute
deviations about the mean statistic; b) The taxicab norm $||{\bf P}%
_{1}||_{\infty \rightarrow 1},$ via (15) and (16), shows that uniform
weights are affected to the columns and the rows; c) The cut norm $4||{\bf P}%
_{1}{\bf ||}_{\boxdot }$ shows that the computation of each principal
dimension of TCA corresponds to balanced 2-blocks seriation, with equality
of the cut norm in the 4 associated blocks.

\medskip
\begin{verbatim}
Acknowledgements.
\end{verbatim}

Choulakian's research has been supported by NSERC grant (RGPIN-2017-05092)
of Canada. The authors thank William Alexander Digout for help in
computations.\bigskip

\section{\bf References}

Beh, E. and Lombardo, R. (2014). {\it Correspondence Analysis: Theory,
Practice and New Strategies}. N.Y: Wiley.

Benz\'{e}cri, J.P. (1973).\ {\it L'Analyse des Donn\'{e}es: Vol. 2:
L'Analyse des Correspondances}. Paris: Dunod.

Benz\'{e}cri, J.P (1992). {\it Correspondence Analysis Handbook}. N.Y:
Marcel Dekker.

Choulakian, V. (2006). Taxicab correspondence analysis. {\it Psychometrika,}
71, 333-345

Choulakian, V. (2008a). Taxicab correspondence analysis of contingency
tables with one heavyweight column. {\it Psychometrika}, 73, 309-319.

Choulakian, V. (2008b). Multiple taxicab correspondence analysis. {\it %
Advances in Data Analysis and Classification}, 2, 177-206.

Choulakian, V. and de Tibeiro, J. (2013a). Graph partitioning by
correspondence analysis and taxicab correspondence analysis. {\it Journal of
Classification}, 30, 397-427.

Choulakian, V., Allard, J. and Simonetti, B. (2013b). Multiple taxicab
correspondence analysis of a survey related to health services. {\it Journal
of Data Science}, 11(2), 205-229.

Choulakian, V., Simonetti, B. and Gia, T.P. (2014). Some further aspects of
taxicab correspondence analysis. {\it Statistical Methods and Applications},
available online.

Choulakian, V. (2016). Matrix factorizations based on induced norms. {\it %
Statistics, Optimization and Information Computing}, 4, 1-14.

Choulakian, V. (2017). Taxicab correspondence analysis of sparse two-way
contingency tables. {\it Statistica Applicata - Italian Journal of Applied
Statistics,} 29 (2-3), 153-179.

Faust, K. (2005). Using correspondence analysis for joint displays of
affiliation networks. In: Carrington, P.J., Scott, J., Wasserman, S. (Eds.),
{\it Models and Methods in Social Network Analysis}. Cambridge University
Press, Cambridge, 117--147.

Gifi, A. (1990). {\it Nonlinear Multivariate Analysis. }N.Y:{\it \ } Wiley.

Gorard, S. (2015). Introducing the mean absolute deviation `effect' size.
{\it International Journal of Research \& Method in Education}, 38 (2),
105--114.

Greenacre, M. (1984). {\it Theory and Applications of Correspondence Analysis%
}. Academic Press, London.

Jensen, S.T. (1999). {\it The Laguerre-Samuelson inequality with extensions
and applications in Statistics and Matrix Theory}. MS thesis, McGill
University.

Khot, S. and Naor, A. (2012). Grothendieck-type inequalities in
combinatorial optimization. {\it Communications on Pure and Applied
Mathematics}, Vol. LXV, 992--1035.

Le Roux, B. and Rouanet, H. (2004). {\it Geometric Data Analysis. From
Correspondence Analysis to Structured Data Analysis}. Dordrecht:
Kluwer--Springer.

Liiv, I. (2010). Seriation and matrix reordering methods: An historical
overview. {\it Statistical Analysis and Data Mining}, 3(2), $70-91$.

Murtagh, F. (2005). {\it Correspondence Analysis and Data Coding with Java
and R}. Boca Raton, FL., Chapman \& Hall/CRC.

Nishisato, S. (1984). Forced classification: A simple application of a
quantification method. {\it Psychometrika}, 49(1), 25-36.

Nishisato, S. (2007). {\it Mutidimensional nonlinear descriptive analysis}.
Chapman \& Hall/CRC, Boca Raton.

Pham-Gia, T. and Hung, T.L. (2001). The mean and median absolute deviations.
{\it Mathematical and Computer Modelling}, 34, 921-936.

Schepers, J., Bock, H-H. and Van Mechelen, I. (2017). Maximal interaction
two-mode clustering. {\it Journal of Classification}, 34, 49-75.

Yitzhaki, S. and Lambert, P.J. (2013). The relationship between the absolute
deviation from a quantile and Gini's mean difference. {\it Metron}, 71,
97--104

\bigskip

{\it Appendix A: List of Western Hemisphere Organizations}

1.\qquad Association of Caribbean States (ACS): Trade group sponsored by the
Caribbean Commnnity and Common Market (CARlCOM).

2.\qquad Latin American Integration Association (ALADI): Free trade
organization.

3.\qquad Amazon Pact: Promotes development of Amazonian territories.

4.\qquad Andean Pact: Promotes development of members through economic and
social integration.

5.\qquad Caribbean Commnnity and Common Market (CARICOM): Caribbean trade
organization; promotes economic development of members.

6.\qquad Group of Latin American and Caribbean Sugar Exporting Countries
(GEPLACEA): Sugar-producing and exporting countries.

7.\qquad Group of Rio: Organization for joint political action.

8.\qquad Group of Three (G-3): Trade organization.

9.\qquad Inter-American Development Bank (IDB): Promotes development of
member nations.

10.\qquad South American Common Market (MERCOSUR): Increases economic
cooperation in the region.

11.\qquad North American Free Trade Agreement (NAFTA): Free trade
organization.

12.\qquad Organization of American States (OAS): Promotes peace, security,
economic, and social development in the Western Hemisphere.

13.\qquad Central American Parliament (PARLAC\'{E}N): Works for the
political integration of Central America.

14.\qquad San Jos\'{e} Group: Promotes regional economic integration.

15.\qquad Latin American Economie System (SELA): Promotes economic and
social development of member nations.

\end{document}